\input{aipcheck}

\documentclass[
    ,final            
  ]
  {aipproc}

\layoutstyle{6x9}

\begin{document}

\title{Seeded quantum FEL at 478 keV}


\author{M. M. G\"unther}{
  address={Max-Planck-Institut f\"ur Quantenoptik, D-85748 Garching, Germany}
}

\author{M. Jentschel}{ 
  address={Institut Laue-Langevin, F-38042 Grenoble}
}

\author{P. G. Thirolf}{
  address={Ludwig-Maximilians-Universit\"at M\"unchen, D-85748 Garching, Germany}
}
\author{T. Seggebrock}{
  address={Ludwig-Maximilians-Universit\"at M\"unchen, D-85748 Garching, Germany}
}

\author{D. Habs}{ 
  address={Ludwig-Maximilians-Universit\"at M\"unchen, D-85748 Garching, Germany}
  ,altaddress={Max-Planck-Institut f\"ur Quantenoptik, D-85748 Garching, Germany}
}

\begin{abstract}
 We present for the first time the concept of a seeded $\gamma$ quantum Free-Electron-Laser (QFEL) at 478 keV, which has very different properties compared to a classical. The basic concept is to produce a highly brilliant $\gamma$ beam via SASE. To produce highly intense and coherent $\gamma$ beam, we intend to use a seeded FEL scheme. Important for the production of such a $\gamma$ beam are novel refractive $\gamma$-lenses for focusing and an efficient monochromator, allowing to generate a very intense and coherent seed beam. The energy of the $\gamma$ beam is 478 keV, corresponding to a wavelength in the sub-\AA ngstr\o m regime (1/38 \AA). To realize a coherent $\gamma$ beam at 478 keV, it is necessary to use a quantum FEL design. At such high radiation energies a classical description of the $\gamma$-FEL becomes wrong.
\end{abstract}

\maketitle


One possibility to produce high energy FEL radiation is to use a conventional SASE FEL design \cite{Saldin2000}. The problem with producing very short wavelengths is that the electron bunch length is much larger than the desired radiation wavelength. The process of microbunching can solve this problem. According to conventional FELs using magnetic undulators, the shortest wavelength to date was produced by the Linac Coherent Light Source (LCLS) at SLAC in Stanford \cite{Emma2010} (1.5 \AA). Another possibillity to produce high energy radiation is to use the new ``asymmetrical'' laser-electron Compton back-scattering scheme (as foreseen for the MEGa-Ray and ELI-NP facillities) presented by C. Barty \cite{ELINP}. In this concept, a long laser pulse (few ns) and very short electron bunches are used. Longer laser pulses means lower bandwidth and therefore lower $\gamma$-radiation bandwidth. Short electron bunches with high repetition rate allow for lower charge per bunch. The consequence is a small electron energy spread and a smaller emittance. Furthermore, the radiation repetition rate can be increased with high beam quality. A contrary concept is to use a laser pulse and electron bunch with the same length (proposed by L. Serafini \cite{ELINP}). But this concept is not reasonable for the production of a high-brilliance $\gamma$ beam. It uses larger electron bunches, which lead to a higher energy spread and a worse emittance. A shorter laser pulse duration leads to a larger bandwidth. The best conditions for the realization of a QFEL is the ``asymmetrical'' Compton back-scattering concept, corresponding to a $\gamma$-FEL with laser wiggler.

For a classical FEL, the main properties are controlled by the dimensionless 3D Pierce parameter $\rho=[\frac{I}{\gamma \cdot I_{A}}(\frac{f_{c}^{2}\cdot K^{2}}{1+K})]^{1/2}$ \cite{Reiche1999}.
Here $I$ is the electron current, $I_{A}=$17 kA the Alfv\'{e}n current, $E_{e}=\gamma \cdot m_{0}c^{2}$ the electron energy, $f_{c}$ the coupling factor and $K$ the undulator parameter which equates to the normalized vector potential $a_{w}$ for a laser wiggler. Important deduced quantities are the gain length $L_{g}=\frac{\lambda_{u}}{4\pi \rho}$ ($\lambda_{u}$ being the undulator period) and the energy spread criterion $\Delta \gamma / \gamma \leq \rho$. A short gain length requires a high bunch current, a small electron beam diameter, a low electron energy spread and a short undulator period. It was shown within the classical FEL model that due to the electron energy diffusion (a first order quantum effect) there is a maximum radiation energy of about 100 keV \cite{Saldin2004, XFEL}, where a FEL can be realized. At higher $\gamma$ energies a classical FEL description becomes wrong. A new quantum FEL description was developed by R. Bonifacio and G. Robb \cite{Bonifacio2009, Bonifacio2011}, which results in very different properties. At the same time it correctly describes the transition to the classical regime. The SASE FEL is no longer chaotic with a spiky spectrum, but emitts a single-frequency sharp line in the inner rest frame of the electron bunch. A new QFEL Pierce parameter $\bar{\rho} =\rho \frac{mc\gamma}{\hbar k}$ is introduced and the quantum regime is reached, if the condition $\bar{\rho} \leq 1$ is fulfilled. The energy spread of the electrons ($\rho mc\gamma$) is smaller than the $\gamma$-radiation energy. The classical cooperation length is deduced by $L_{c}=\frac{\lambda_{r}}{4\pi \rho}$ ($\lambda_{r}$: radiation wavelength). In the quantum regime the cooperation length $L'_{c}=\frac{L_{c}}{\sqrt{\bar{\rho}}}$ is much increased and the requirement for the electron beam energy spread is $\Delta \gamma/\gamma \leq 4\rho \sqrt{\bar{\rho}}$ \cite{Piovella2006}. Furtheremore, the gain length in the quantum regime is given by $L'_{g}=\frac{\lambda_{L}}{8\pi \rho\sqrt{\bar{\rho}}}(\sqrt{1+\bar{\rho}})$. An important basis for a QFEL is a very good electron beam with a low normalized emittance and a very small energy spread. The newly predicted quantum properties of the electron beam - laser-wiggler should already be observable at the MEGa-Ray facillity.

\begin{figure}
  \includegraphics[width=0.47 \textwidth]{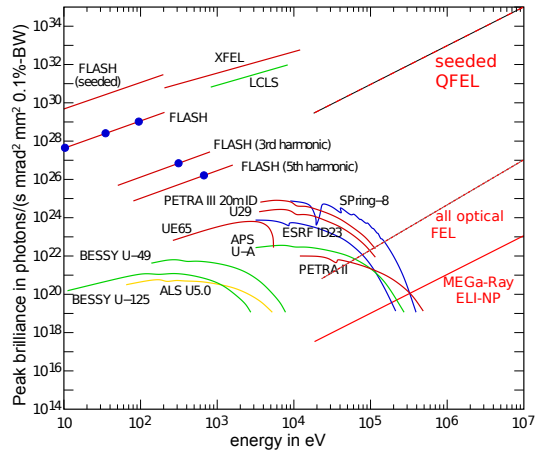}
  \caption{Peak brilliance as a function of radiation energy. The peak brilliance of the seeded QFEL radiation is several orders of magnitude higher than the SASE FEL radiation.}
  \label{fig:peakbrilliance}
\end{figure}

\begin{figure}
  \includegraphics[width=\textwidth]{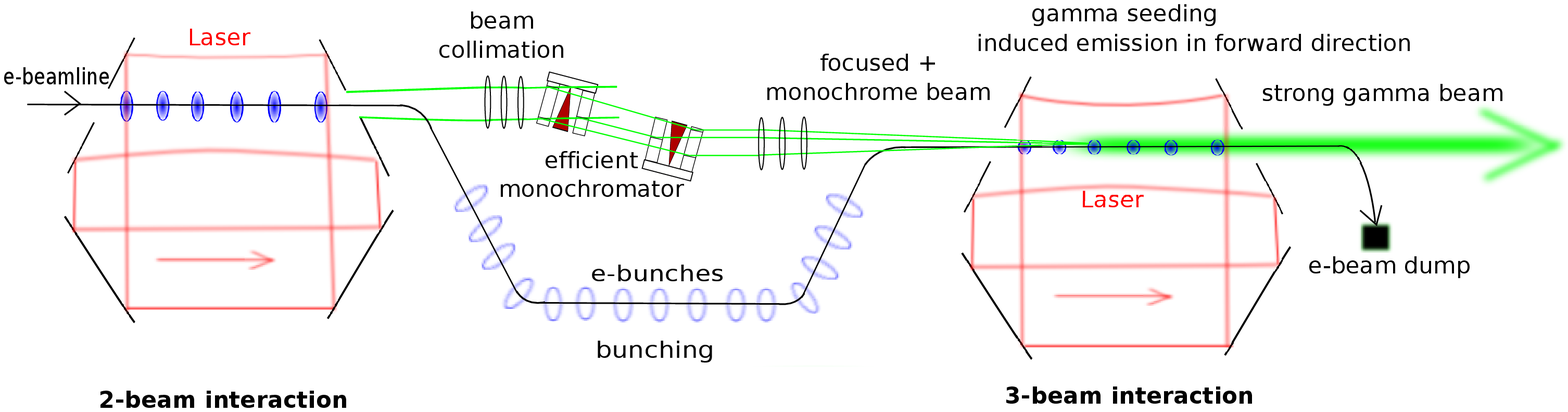}
  \caption{Schematics of a seeded $\gamma$-FEL.}
  \label{fig:gammafel}
\end{figure}

In the following we will describe a seeded $\gamma$-QFEL concept for 478 keV. The important parameters are shown in Tab. \ref{tab:parameter}. According to R. Bonifacio et al. \cite{Bonifacio2005}, it is possible to estimate the parameters for a QFEL with a laser wiggler. Then, the relations between the radiation and the wiggler in the quantum regime are given by the parameter $\eta$, which contains the wiggler laser wavelength $\lambda_{L}$ in units of $\mu$m and the radiation wavelength $\lambda_{r}$ in units of \AA . $\eta$ is determined by the relation $\lambda_{r}($\AA$)=\frac{\lambda^{3}_{L}(\mu m)}{\eta^{2}}(1+a^{2}_{w})$, which is fulfilled for $\eta=$6.3. Furthermore, $\eta=\epsilon_{1}\epsilon_{2}\epsilon_{n}$(mm mrad), where $\epsilon_{n}$ is the normalized emittance of the electron beam in units of mm mrad. $\epsilon_{1}$ as well as $\epsilon_{2}$ are conditions for the QFEL and combine the laser wiggler parameter with the electron beam parameter: $\epsilon_{1}=\frac{W_{0}}{4\sigma_{0}}\geq 1$ and $\epsilon_{2}=\frac{\beta^{*}}{Z_{0}}\geq 1$ ($W_{0}$: laser waist, $\sigma_{0}$: minimum electron beam diameter, $\beta^{*}$: Rayleigh-length of the electron beam, $Z_{0}$: laser Rayleigh-length). For a QFEL at 478 keV the conditions are complied with $\epsilon_{1}=\epsilon_{2}=4.4$. In addition, we can match the wiggler and the electron beam parameter in the quantum regime considering the relations $Z_{0}=\frac{\pi W_{0}}{\lambda_{L}}$ and $\sigma_{0}=\sqrt{\frac{\epsilon_{n}\beta^{*}}{\gamma}}$. The results are shown in Tab. \ref{tab:parameter}. The SASE QFEL can produce radiation with a very small line width. However, the radiation emission of the photons in the lab system will be several hundreds of $\mu$rad. A seeded QFEL concept using novel $\gamma$ optics enforces very narrow emission in the lab system. Figure \ref{fig:peakbrilliance} shows the peak brilliance of $\gamma$ and X-ray radiation facilities. The estimated peak brilliance of a seeded QFEL is about 8 orders of magnitude higher than for an all optical FEL. A seed is produced by compton back-scattering like MEGa-Ray and then is monochromatized, collimated and focused by $\gamma$ optics. This seed beam interacts within a second laser wiggler with electron bunches. Figure \ref{fig:gammafel} shows a scheme of a seeded $\gamma$ FEL for 478 keV. In the first stage a seed beam is produced by Compton back-scattering off an intense laser-wiggler pulse at a string of high energy electron bunches accelerated by X-band technology. After the first stage, the electron bunch train is deflected in a chicane, while the produced $\gamma$ seed beam is collimated and monochromatized using novel $\gamma$ optics \cite{Habs2011, Jentschel2011}. The $\gamma$ optics part is very important for the realization of a seeded $\gamma$ FEL. Using a further $\gamma$ lens system, the seed beam is focused into a second laser-wiggler stage. The three-beam interaction stage produces a very intense and coherent $\gamma$ beam.
 In the hard X-ray range the index of refraction is around 1. However, there is an energy dependence of the index of refraction which leads to a small deviation $\delta$ from 1. A recent experiment at the Institut Laue-Langevin (ILL) in Grenoble has shown that above a photon energy of 700 keV the sign of $\delta$ is positive in silicon with an enhancement around about 1 MeV \cite{Habs2011}. A positive $\delta$ corresponds to use of convex lenses for focusing. In case of high Z materials, we estimate a positive delta already below 700 keV. In addition, we can use novel $\gamma$ optics with a high index of refraction for 478 keV to match the seed beam efficiently. According to the electron beam parameters (Tab. \ref{tab:parameter}), the angle of divergence of the seed beam is arround 500 $\mu$rad. Using $\gamma$ lenses, the beam can be further parallelized to 5 $\mu$rad while keeping the emittance. A flat-crystal monochromator can be used for monochromatization. Increasing the efficiency of the flat-crystal monochromator, we intend to use a multiple angle shifter in the double flat-crystal monochromator to extract many angular bins for the same $\gamma$ energy, making the monochromator much more efficient \cite{Jentschel2011}. In addition, the intensity can be enhanced by about a factor 10$^{4}$. Furthermore, using $\gamma$ lenses and an efficient monochromator, the beam gains partially transversal coherence as well as partially longitudinal coherence.
\begin{table}
\begin{tabular}{ll|l|ll|l|}
\hline
electron beam& &laser wiggler &seed beam& &FEL parameters\\
\hline
charge & 20 pC & $\lambda_{L}=1$ $\mu$m & $\lambda_{r}$ & 1/38 \AA & $\rho=9\times 10^{-5}$ \\
duration & 0.5 ps & $a_{w}=0.03$ & bandwidth & 10$^{-6}$& $\bar{\rho}=2.6\times 10^{-2}$\\
current & 40 A & $Z_{0}=24$ mm & $\gamma$'s/microbunch & 300& $L_{g}=0.8$ mm\\
Rep. rate & 120 Hz & $W_{0}=88$ $\mu$m & spot size & few nm& $L'_{g}=0.4$ mm \\
$\sigma$ & 50-5 $\mu$m & $P_{w}=7.5$ TW & &&$L_{c}=21$ \AA \\
$E_{e}$ & 125 MeV &&&&$L'_{c}=130$ \AA \\
$\Delta \gamma / \gamma$ & $\leq$10$^{-3}$-10$^{-4}$ &&&&\\
$\epsilon_{n}$ & 0.075 mm mrad &&&&\\
$\beta^{*}$ & 10 cm &&&&\\
\hline
\end{tabular}
\caption{Parameter of a seeded QFEL concept.}
\label{tab:parameter}
\end{table}
The seed beam causes within a second stage an induced emission in forward direction. The result is a very brilliant $\gamma$ beam. The maximum expected spectral flux is 10$^{11}$ eV$^{-1}$s$^{-1}$.

Not only the peak brilliance can be increased. For the first time a coherent $\gamma$ beam can be produced with $N_{\gamma}\approx 10^7$ coherent photons in a phase-space cell, which results in a much larger nonlinear coupling $N_{\gamma} \alpha$ with charges of hadrons, leptons and other particles. A $\gamma$ QFEL can open up totally new areas of fundamental physics and applications \cite{Paganin2006}. Exciting nuclear resonances slightly off resonance, one gets very large positive or negative indices of refraction, thus enabling an isotope-specific diffraction. The 478 keV M1 transition of $^{7}$Li can be produced by a $\gamma$ QFEL, which is attractive for investigations in "green energy" research such as Li-battery optimization by $\mu$m tomography. In life-science research, the detection of Li deposition in the brain for manic-depressive psychosis treatment can be imaged with a high spatial resolution.


  



\bibliographystyle{aipproc}   





\end{document}